# New insights for setting up contractual options for demand side flexibility

Gianfranco Chicco, Andrea Mazza
Politecnico di Torino, Dipartimento Energia "Galileo Ferraris"
Corso Duca degli Abruzzi 24, 10129 Torino, Italy
gianfranco.chicco@polito.it, andrea.mazza@polito.it

**Abstract**
This paper exploits the Duration-of-Use of the demand patterns as a key concept for dealing with demand side flexibility. Starting from the consideration that fine-grained energy metering is not used at the point of supply of the electricity consumers, i.e., the granularity of the energy measured (at time steps of 15 minutes, 30 minutes or one hour), the event-based energy metering (EDM) is indicated as a viable option to provides a very detailed reconstruction of the demand patterns. The use of EDM enables high-quality tracking of the demand peaks with a reduced number of data with respect to the ones needed to measure energy at regular time steps for reaching a similar peak tracking capability. From the EDM outcomes, a new class of options for setting up tariffs or contracts for flexibility, based on the demand duration curve, is envisioned.

## 1. Introduction

After the restructuring of the electricity business, with the beginning of the competitive electricity markets and the unbundling of the generation, transmission, distribution and retail services, the regulation has changed. Grid codes have been established for the networks, and performance-based ratemaking (PBR) [1] has been applied in different contexts. With PBR (also called performance-based regulation [2]), cost reductions and increased efficiency in the energy management may lead to better profits for the electric companies. PBR is the alternative to the traditional cost-of-service (CoS) based regulation, and the profits are obtained from meeting the objectives for delivering a reliable, affordable and clean power and energy system. To avoid potential cost reductions that can affect the quality of service, quality control is indeed needed by the action of the regulators [3]. According with [4], a well-designed scheme for PBR could "drive important societal outcomes, as well as create new business opportunities for innovative utilities and third-party players". In the U.S., many utilities are looking for a change towards PBR. In the Utility Dive's 2018 State of the Electric Utility survey [5], based on an online questionnaire sent to Utility Dive readers in December 2017, only 8% of utility respondents indicated to want CoS regulation, while 44% indicated preference for a hybrid model that mixes traditional CoS with PBR, and 32% are favourable to a model with predominant PBR.

In Europe, in the last two decades the quality of supply has been subject to new types of limits, defined in terms of non-exceeding probabilities on relevant quantities for given time periods, rather than using a simple threshold for these quantities. An example is the quality of supply, with the regulation established by the European Standard EN 50160. In this Standard, for harmonic voltages it is established that in normal operating conditions, during each period of the week, 95% of the RMS voltage values averaged in 10 minutes for each individual harmonic voltage have to be lower or equal to the value indicated in the corresponding table in the Standard. To assess the relevant quantities, fine-grained monitoring is needed. Each quantity is calculated for a base interval of 200 ms, then the results are first aggregated over very short-term intervals (3 seconds) by calculating the Euclidean average of the quantity for 15 successive base intervals. A successive aggregation over short-term intervals (10 minutes) is determined by calculating the Euclidean average of the quantity for 200 successive very short-term intervals. Finally, aggregation over long-term intervals (2 hours) is obtained from the Euclidean average of the quantity for 12 successive short-term intervals.

The approach based on per cent values indicated above has not been applied yet for energy management purposes, mainly because of lack of detailed representation of the demand patterns. The traditional energy metering provides results for time steps of 15 minutes, 30 minutes or one hour. However, as discussed in Section 4.2, these time steps are too long to provide useful information for flexibility purposes. Using smart meters that gather data regularly at shorter time steps from several energy meters imposes a high burden on the communication channels. In addition, the use of constant time steps is inefficient to identify the actual peaks and ramps appearing in the demand patterns. This has also a remarkable effect on the data available to assess the operation of the low voltage (LV) grid, and in the determination of quantities relevant to the grid efficiency (e.g., the network losses, which depend on the actual power flows in a non-linear way, and may be incorrectly assessed if the power is not determined properly). In [6] it is indicated that "*the fact that the LV grid cannot provide sufficient data today, will have also a negative impact on the future LV grid investment strategy*", and also that lack of sufficient data to enable acceptable grid observability does not only hinder grid monitoring, but also the objective of obtaining an affordable energy system. The recent manufacturing of an event-driven energy meter [7] based on the generation of events linked to the characteristics of the demand patterns is opening new paths to the definition of innovative types of tariff options, based on the concept of Duration-of-Use. In this way, flexibility can be associated to changes occurring in the duration curve drawn from the demand pattern.

This paper discusses the possibility of enhancing flexibility on the demand side through the provision of data that better reflect the characteristic of the demand patterns. In particular, the use of a new type of meter, based on measuring energy demand between events generated in asynchronous mode, is shown to be very promising to reconstruct demand patterns with much better precision than in the traditional interval metering.

Specific advantages related to the use of EDM are discussed in the identification of the demand peak and in the possible reduction of the data collected by the meter. Further positive aspects of the better knowledge of the demand patterns are the possibility to estimate the technical losses occurring in the supply grid, especially when the demand has large variations, as it happens in the last mile of the electricity distribution.

The next sections of this paper are organised as follows. Section 2 illustrates the evolution of the concepts referring to demand side response and demand side flexibility. Section 3 recalls the characteristics of event-driven energy metering. Section 4 focuses on how the EDM outcomes provide new insights for flexibility assessment, and contains a worked example that uses the demand pattern of an individual consumer, whose results prelude to the illustration of the proposed Duration-of-Use scheme for electricity pricing in Section 5. The last section contains the conclusions.

## 2. Demand side response and demand side flexibility

The evolution of rulemaking has considered for many years demand side response (DSR). More recently, the terminology has been shifted to demand side flexibility (DSF), even though some of the main concepts indicated in the regulatory documents still refer to DSR. However, flexibility means more than DSR. The Council of European Energy Regulators (CEER) in the Conclusion paper [8] indicates that "*Flexibility is the capacity of the electricity system to respond to changes that may affect the balance of supply and demand at all times*". The European Smart Grids Task Force Expert Group 3 [6] defines flexibility as "*the ability of a customer (Prosumer) to deviate from its normal electricity consumption (production) profile, in response to price signals or market incentives*".

The concept of *flexibility*, applied to the demand side, may be compared with the same concept seen at the generation side and, on this basis, a common understanding should be considered. As reported in [9], flexibility is first of all "responsiveness", that is not only energy or power (capacity), but requires their provision with a well-defined shape. This is exactly what is used for the supply side: the flexibility of the traditional generators indicates the capability of them to "ramp up" or "ramp down" to reach the desired operation condition. An example of normalized flexibility index referring to the supply side incorporating information on ramp rates for supply side is reported in [10]. However, due to the fact that the demand flexibility involves consumers, a non-negligible aspect is how many of them would like to participate at the program, because making unrealistic assumptions about their engagement can compromise its effectiveness [11]. Furthermore, smart meters with appropriate characteristics are needed to allow better deployment of DSF options, as also confirmed by [12], stating that "*Smart Meters are a fundamental enabling factor for DSF, however some of the current generation of smart meters may not be adequate to deliver all services required for DSF and standardisation across Europe remains an issue*". In [8] this concept is reinforced by the results of a survey conducted by CEER about the possibility of incentivising consumers to use the network in the most efficient way. From these results, most of the respondents considered that network tariffs cannot have sufficient granularity to send price signals corresponding to the exact local flexibility need. Moreover, in [8] the main constraints indicated to the practical feasibility of a real-time market are lack of liquidity and lack of technical tools (smart-metering, hourly measurement, real time monitoring). In this respect, advanced metering may enable the definition of more refined tariff structures and options. Smart Metering, Regulatory Framework for Tariff Structures and Contractual Arrangements are considered as the most effective means for achieving flexibility use at distribution level. In [6] there is an explicit recommendation to the European Union to create a smart meter roadmap, in which smart meters should be able to meet the requirements of future markets (e.g., high-resolution time intervals), providing a modular and flexible architecture for the metering infrastructure. Furthermore, it is indicated that some characteristics (such as the minimum size of the product, or the temporal granularity) should be identified to allow the contribution of all resources to the provision of the needed services.

The demand shifting introduced by DSF has an impact on multiple aspects:
- Reduction of the demand peaks, with the benefit of reducing the need for running expensive "peaking" generation to cover the peaks. Peak reduction also helps in reducing grid losses and congestion in the grid in time periods with high power flows.
- Reduction of the curtailment of intermittent generation (e.g., from wind and photovoltaic plants),
- Postponing investments on new network assets by reducing the peak net demand (i.e., local demand minus local generation, also taking into account the charge or discharge of local storage units) seen from the point of connection with the grid.
- Increasing the system efficiency, also by making it possible the operation of the equipment at higher efficiency, also leading to environmental benefits.
- Increasing reliability, avoiding the need for installing generation capacity that could be used rarely, or avoiding load shedding or load curtailment.

In addition, DSF may provide reserve capacity to the system, available at different time scales after request, thus reducing the need for procuring part of the reserves from traditional generation. The time scales at which DSF may operate are in general even faster than the one referring to conventional generation, making DSF highly valuable. The position paper [13] considers the principles to ensure that consumers are able to offer their flexibility, indicating that consumers "*should at least have the choice to be metered and settled at the same time resolution as the imbalance period in national markets when it is technically possible*".

Following the literature regarding DSR, the consumers could offer their flexibility according to price-based or incentive-based schemes [11]. By comparing the definitions reported in technical documents, the flexibility that can be obtained from price-based scheme is known as *implicit flexibility*, whereas the one obtained through incentive-based schemes is called *explicit flexibility* [14].

With implicit flexibility, the flexibility is delivered according to the sensitivity of the customer to the price signal for time-of-use tariffs (ToU), critical peak pricing (CPP), peak time rebates (PTR), or real time pricing (RTP). ToU tariffs are set in advance (e.g., every year), with time periods subject to the same price that can change during the year, e.g., seasonally. CPP is a version in which higher prices are defined (again, in advance, in general at least a few hours before the expected critical event) for specific time periods in which the network loading may become high and thus critical. In [15] it is indicated that implicit flexibility from ToU or CPP does not reflect the actual conditions. In PTR, the consumers receive reductions in their electricity bill during peak periods established a priori, if they reduce their consumption with respect to a given baseline. However, the definition of the baseline may be critical, because of possible strategic behaviour of the consumers (see below). RTP refers to the operation in normal conditions, where the price changes during time with a short notice (e.g., a few hours) throughout the year. However, the diffusion of RTP as a dynamic pricing option is somehow limited at the moment, as various consumers (e.g., residential and small enterprises) have no access to these tariffs [16].

Conversely, with explicit flexibility, the flexibility is committed by participating in incentive-based programmes that rely upon direct load control, interruptible or curtailable rates, emergency demand response, capacity market, and ancillary service markets. These programmes may be managed by the supplier or by another entity (e.g., an aggregator). The incentive-based framework may be more effective if it provides benefits to the incentive manager that participates in energy markets, capacity markets, of balancing markets. In the impact assessment results shown in [16], commercial and industrial consumers are considered to be likely to participate in incentive-based DSR only through aggregators, while the participation of retail consumers is not expected.

The smart metering system has the objective to ensure that appropriate functionalities and interoperability are available to the consumers. With the availability of new generations of smart meters, there will be room for defining a wider range of DSF offers, facilitating higher competition in the provision of smart energy services. For smart meter deployment, the net benefit has to be assessed as the estimated savings in generation and network capacity minus the costs of meters and activation [16].

The consumer (or prosumer) may decide to deliver flexibility in a voluntary way, unless a contractual arrangement has been established (in this case the provision of flexibility becomes mandatory *de facto*; this arrangement is typical of incentive-based schemes). In [6] it is pointed out that the participation of the demand side in providing different services could be enhanced by allowing *value stacking*, that is, the possibility of offering different services at the same time. Of course, this possibility requires clear determination of the allocation of the DSF provided to the different services.

One of the main aspects to be considered for understanding the impact of the demand side on the system is the creation of demand *baseline*. Its appropriate calculation is fundamental for evaluating the impact of the demand flexibility in the system. As shown in [17], the methodologies for defining the baselines for DSR can be based on appropriate functions. For example, the *initial* baseline is calculated as the *average* demand among the X highest energy usage days out of the prior Y non-event days (this mode is indicated as HighXofY [18]; alternative modes are LowXofY or MidXofY), or with exponential moving average or regression [19]. Another possibility is the use of a control group, which has to be as similar as possible to the remaining population, to well represent the behaviour of the population. Then, the *adjusted* baseline is the adaptation of an initial baseline to the actual load pattern occurring before starting the DSR action. Specific rules have to be followed to determine the adjusted baseline. For example, an adjustment factor is calculated as the difference between the observed demand and the initial baseline for a calibration period starting two hours before the event notification, with a minimum adjustment of zero, and is then applied to modify the initial baseline to get the reference baseline for DSR. The starting time of the calibration period cannot be too early, in order to avoid strategic behaviour of the consumers to artificially create favourable conditions in the determination of the baseline, leading to economic advantages in the determination of the reward after a DSR event [20]. A well-established way to define baselines for flexibility is still needed. In [6], there is a recommendation to categorise the best practices that can be identified for baseline design and validation, possibly referring to the specific flexibility resources. Metering equipment are also required to be able to verify that the load variation is achieved [16].

**3. Event-driven energy metering**

Event-driven energy metering (EDM), also called event-based energy metering, is the acquisition of energy data from measurements gathered at non-regular time steps, triggered by the occurrence of an event. EDM introduces a different paradigm with respect to the classical timer-based metering or timer-driven metering (TDM, also called interval metering), in which energy data are taken from measurements gathered at regular time steps.

The EDM principles have been presented in a number of recent publications, among which [7][21]-[24]. The main characteristics are summarised below.

EDM considers an *elementary time interval* $\tau$ as the shortest duration of time of interest for the representation of the demand pattern. An example used in many of the publications indicated above is $\tau = 1$ s. The elementary time interval is not linked to the internal characteristics of the meter, in which data sampling may occur at much faster sampling rates. The elementary time interval in general should not be too low, to avoid the effects of very fast variations that occur in the energy demand process but can be considered as poorly relevant for the energy-based representation of the demand pattern. Inside the elementary time interval, the demand pattern is represented by using the average power obtained by dividing the energy measured in the elementary time and the duration of the elementary time itself. Faster demand dynamics are then intentionally filtered out in this representation.

EDM is based on a *target* evolution in time of the demand pattern, and generates an *event* each time one of the following conditions occurs:
  a) *change-of-value*: the change of the average power with respect to the previous elementary time interval is higher than a user-defined threshold $\delta_1$; or,
  b) *accumulated energy variation*: the variations of the average power occurring at successive elementary time intervals are accumulated during time; the event is generated when the sum of these variations exceeds a user-defined threshold $\delta_2$.

While the rationale of the change-of-value is quite intuitive, and has been used in the mechanisms denoted as change-and-transmit in [25], the accumulated energy variation has been introduced in [21] to encompass cases in which the deviation with respect to the target may occur from progressive (even individually small) variations that move the pattern away from the target. Indeed, the combined use of the thresholds $\delta_1$ and $\delta_2$ is the major reason of success of EDM.

The target is modified after the occurrence of each event, to take into account the possible occurrence of a new process with different characteristics. The thresholds $\delta_1$ and $\delta_2$ are imposed in the EDM, and may be adapted during time to provide different views on the demand process.

The detection of an event is followed by the generation of the corresponding *time stamp* information and the recording of the *energy* measured between the previous event and the new one. If the pattern follows the expected trend during time, no data recording is needed. In case of a failure, the pattern exhibits a change that is detected as an event. Consistency with fiscal metering is obtained by defining the end of the billing period as an event.

## 4. EDM provides new insights for flexibility assessment

*4.1. Basic aspects*

The EDM outcomes provide high-valued knowledge of the processes that characterise the energy usage. A major advantage of EDM is the possibility of reconstructing the demand pattern with high detail starting from the information provided from a few events. In fact, the demand pattern between two events is represented by a constant average power. This representation filters out the small demand variations occurring between the two events, but keeps the *exact* information on the actual energy used. In this way, EDM enables high-quality tracking of the demand peaks with a reduced number of data with respect to the ones needed to measure energy at regular time steps for reaching a similar peak tracking capability. In particular, the identification of a peak detected in a single elementary time interval would be possible with TDM only by reducing the TDM time step to the elementary time interval. The effectiveness of the demand pattern reconstruction from TDM or EDM may be quantified by using appropriate metrics [26]. From these metrics, in typical applications to energy systems the better effectiveness of the demand pattern reconstruction from EDM is quite evident. Furthermore, EDM provides information close to real time, including the possibility of obtaining an estimate of the expected demand if no new event occurs, providing some anticipatory knowledge with known uncertainty. In fact, in the absence of new events, at each elementary time interval the demand changes within the limits set by the threshold $\delta_1$. These aspects are also fully relevant to enhance the role of metering in applications targeted at the needs of Industry 4.0 [27].

The use of EDM enables significant improvements with respect to timer-based electricity pricing structures that may induce the birth of new peaks due to the synchronisation of the demand. This effect is known in electricity markets: if there is a stepwise reduction of the price known in advance, the consumers may respond with programming higher usage of their devices immediately after the price variation, thus creating a loss of diversity in the natural usage of the devices (that is typically not synchronised). This causes a *payback* or *rebound* peak in the individual demand, that also reflects on the aggregate demand. To reduce this effect, the use of Multi-ToU and Multi-CPP electricity pricing has been proposed in [28]. In the Multi-ToU scheme, the consumers are partitioned into groups, and different ToU prices are applied to each group, in such a way to avoid synchronisation of the price changes. The cost function is determined in such a way that the expenditure of the consumer is minimised. In this way, the consumers receive a kind of compensation for providing flexibility. In the results shown, the Multi-ToU solution has the effect of flattening the residential demand. The Multi-CPP scheme achieves a similar result for the time periods corresponding to emergency events. With EDM, there is no fixed timing, and the demand peaks can be identified in an accurate way. As such, it is easy to formulate tariff structures that avoid the rebound peak because are based on the duration curve rather than on the demand pattern (these aspects are discussed in Section 5).

*4.2. Worked example*

The concepts illustrated in the previous sections are applied to the demand pattern of a residential consumer [29], whose data are available for the time step of 1 s and for a period of observation of one day, i.e., 86400 s (Figure 1). These initial data (that in practice are not visible from the metering outcomes, but are assumed to be known here in order to make specific considerations on the methods used) are processed by considering the elementary time interval $\tau = 1$ s for EDM (with different choices for the two thresholds), and by using TDM at time steps of 1 hour, 30 minutes, 15 minutes, and 1 minute (the latter value has been chosen for comparison purposes, even though it is not a typical time step used in the current energy metering practices at the point of connection to the grid).

The demand pattern has multiple peaks and is particularly challenging for testing the EDM effectiveness. The total energy consumption during the day is 7.52 kWh. The peak value of the average power determined at each second is 3574 W.

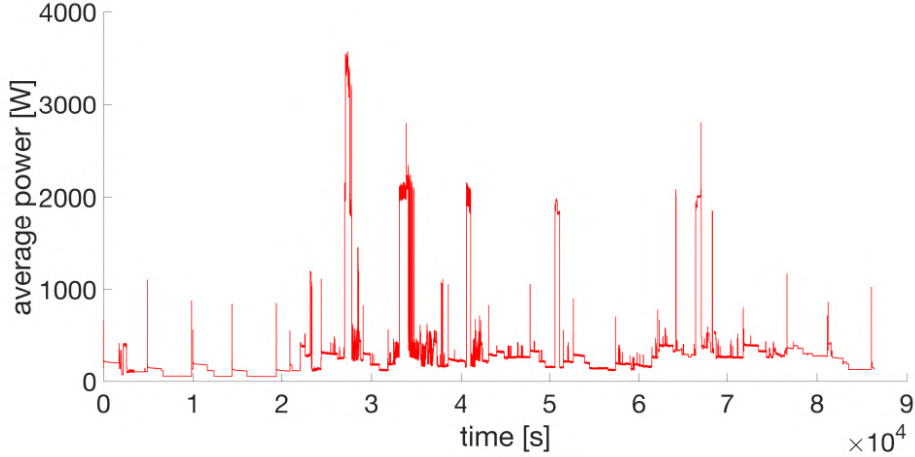

Figure 1. Initial data gathered at 1 s time step.

The results of application of the TDM and EDM representations are shown in Table 1. Clearly, the numerical values depend on the data, and the results cannot be used for reaching conclusions of general validity. In particular, the data used contain more peaks than progressive variations. For this reason, the EDM representation obtained by using the thresholds $\delta_1 = 500$ W and $\delta_2 = 500$ Ws leads to 121 events, of which 108 are triggered by $\delta_1$ (change-of-value type), and 117 are triggered by $\delta_2$ (accumulated energy variations type). From these numbers, it is apparent that there are many events generated by the simultaneous activation of the two thresholds. Likewise, the EDM representation obtained by using the thresholds $\delta_1 = 120$ W and $\delta_2 = 500$ Ws leads to 517 events, of which 517 (i.e., all) are triggered by $\delta_1$, and 99 are triggered (also) by $\delta_2$.

Table 1. Outcomes of different representations for reconstructing the demand pattern.

| Representation | Number of points | Peak average power [W] (and per cent of the peak at 1 s) | Euclidean distance after pattern reconstruction | Daily energy losses (per cent of the losses calculated at 1 s) |
|---|---|---|---|---|
| TDM 60 min | 24 | 932 (26.1%) | 366.3 | 53% |
| TDM 30 min | 48 | 1512 (42.3%) | 326.3 | 63% |
| TDM 15 min | 96 | 2572 (72.0%) | 245.5 | 79% |
| TDM 1 min | 1440 | 3506 (98.1%) | 98.1 | 97% |
| EDM ($\delta_1 = 500$ W, $\delta_2 = 500$ Ws) | 121 | 3428 (95.9%) | 73.6 | 98% |
| EDM ($\delta_1 = 120$ W, $\delta_2 = 500$ Ws) | 517 | 3537 (99.0%) | 46.9 | 99% |

From Table 1 it becomes quantitatively clear that the use of TDM with time step 15 min, 30 min and 60 min provides very poor identification of the actual peaks and of the shape of the demand pattern. Only a TDM with 1 min time step would be effective with these data. Conversely, EDM provides very interesting outcomes by using a lower number of points (events). From the calculation of the Euclidean distances, it is also clear that the pattern reconstructed through EDM has lower distances from the initial data with respect to TDM, thus superior capabilities concerning the reconstruction of the overall demand pattern (not only the peak). To further view this result, Figure 2 shows the reconstructed demand patterns from the four TDM cases and the two EDM cases. In the EDM solution with $\delta_1 = 120$ W and $\delta_2 = 500$ Ws, the number of events (517) becomes relatively higher, however it is still lower than the 1440 points with which TDM cannot reach comparable results. In the last column of Table 1, the daily energy losses on the conductor that supplies the load (in the hypothesis of a cable with 50 m length) are shown in per cent of the daily energy losses occurring in a reference case determined from the initial data at 1 s time step. It appears that the daily energy losses may be determined in a satisfactory way by using the EDM results with $\delta_1$ = 120 W and $\delta_2$ = 500 Ws (99%), in a reasonable way by using the EDM results with $\delta_1$ = 500 W and $\delta_2$ = 500 Ws (98%) and by using the TDM results with 1 min time step (97%), while in the other TDM cases the determination of the daily energy losses is rather far from the value of the reference case. The use of the results coming from TDM with 60 min time step leads to determine only 53% of the daily energy losses. These results explain another strong point to promote EDM as a very effective solution to improve observability of the network-related variables. The improvement of the calculation of the network losses has also important economic implications, as the estimation of the costs of the losses is a key point for the distribution system operators.

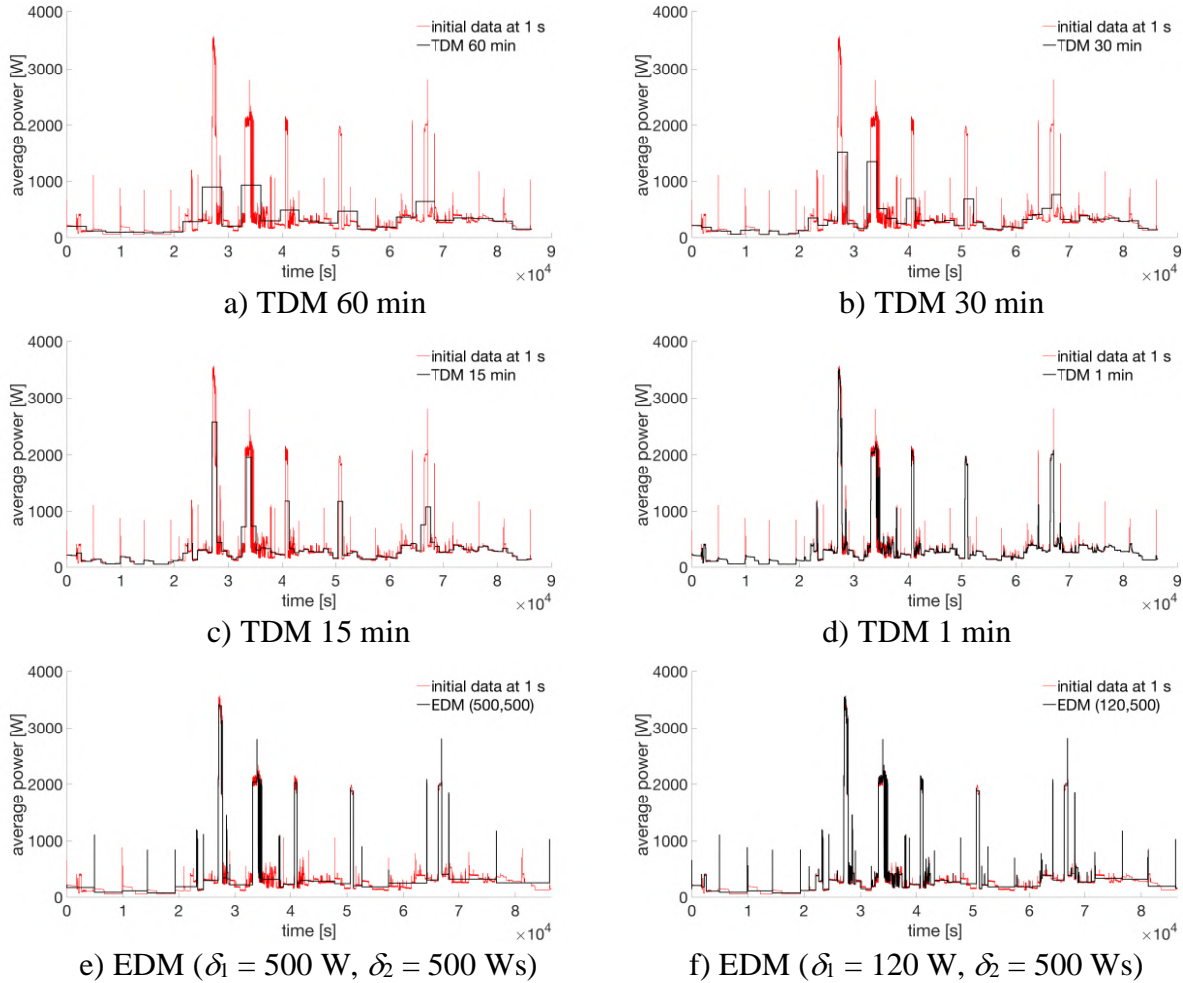

a) TDM 60 min    b) TDM 30 min

c) TDM 15 min    d) TDM 1 min

e) EDM ($\delta_1$ = 500 W, $\delta_2$ = 500 Ws)    f) EDM ($\delta_1$ = 120 W, $\delta_2$ = 500 Ws)

Figure 2. Demand pattern reconstruction from TDM and EDM outcomes.

## 5. Duration-of-Use scheme for electricity pricing

*5.1. Underlying principles for a Duration-of-Use pricing scheme*

The basics aspects to describe the Duration-of-Use (DoU) scheme come from the analysis of Figure 3. The duration curve is constructed by simply *sorting* all the entries of the same demand pattern (reconstructed at 1 s for the sake of comparison) in descending order. The full duration curve of Figure 3a indicates how the main differences refer to the representation of the peaks. The zoom of Figure 3b provides a more detailed picture of the relations among the different representations of the reconstructed demand pattern and the reference pattern with data gathered at 1 s.

The first remark is that hourly-based data gathering (as done today in many real systems) is totally ineffective for the purpose of considering the demand peaks. If a TDM scheme should be used, it should indicatively operate with time steps not higher than 1 minute. In fact, with respect to the peak detected from

TDM at 1 minute (3506 W), any peak pricing scheme constructed on the hourly data for peaks higher than 932 W would find no action needed on the demand pattern to reduce the peak. This is clearly a misleading outcome, as the existing peaks are hidden in the representation based on hourly metered data.

Thereby, the time step of 1 minute seems a reasonable option for interval metering. However, instead of using real-time pricing with minute pricing (for which prerequisite would be the installation of smart meters with 1-minute time step), event-driven energy metering provides a very detailed reconstruction of the demand patterns – numerical evidence has been shown in Section 4, by using less data and opening the possibility of applying the innovative DoU scheme for electricity pricing to enhance the options to provide flexibility from the demand side.

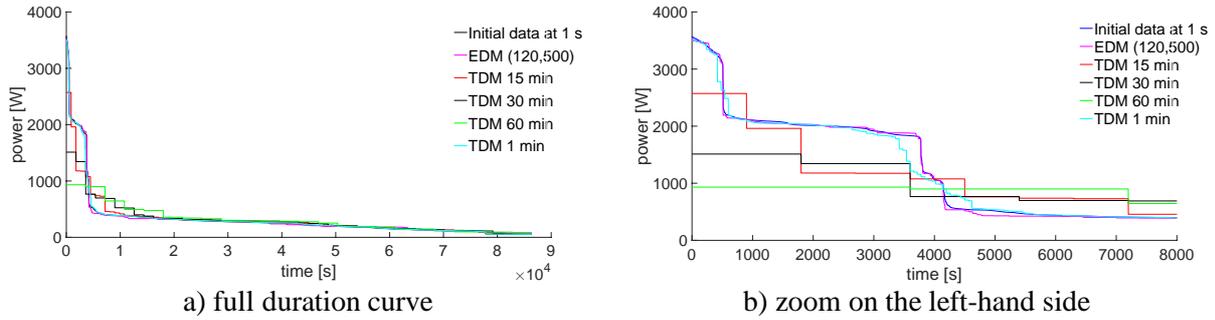

a) full duration curve    b) zoom on the left-hand side

Figure 3. Duration curves from TDM and EDM outcomes.

The proposed DoU scheme is based on setting up limits onto the demand duration curve. On the time axis, the limits are multiple of the EDM elementary time interval, and are progressively extended up to a given portion of the total time observed $T_x$. A qualitative example is presented in Figure 4, in which there are two time limits defined on the duration curve, namely, $T_{d1}$ and $T_{d2}$, with the corresponding power values $P_{d1}$ and $P_{d2}$, plus the limit $P_{d0}$ valid until the time $T_x$.

Some remarks on the example presented in Figure 4:
- The time horizon $T_x$ depends on what the operator considers to be relevant for demand flexibility purposes (e.g., one day, or shorter time intervals).
- The representation of the limits is stepwise (for the sake of easier representation), however in general it may be any function defined by the operator in charge of setting the relevant option.
- The limits defined may be fixed, or change periodically, even from one time interval to another, to reproduce the same dynamics already existing from ToU to RTP. In addition, these limits may be set up in absolute values of power, or in per cent values of a reference power (e.g., the contract power).
- The area below the limits also defines the maximum energy to be consumed in the observation period.

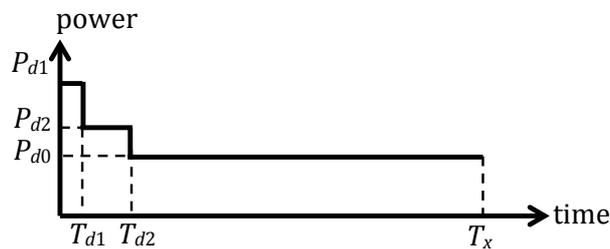

Figure 4. Conceptual sketch of the limits imposed by the DoU principle.

Once the limits have been defined, how is it possible to construct a scheme for incentivising (or penalising) the consumers? Indeed, there are different possibilities, as indicated in many PBR practices. In general, the demand duration curve originated by the EDM results is compared with the limits, to assess the amount of time, power and energy that exceed the prescribed limits. Then, the exceeding quantities may be penalised, with the objective of reaching a flatter demand pattern, leaving the consumer the choice on how to reach this objective. The latter point is quite important, because the absence of a defined time slot with better conditions avoids to induce synchronised peaks in the modified demand pattern, and contributes to the possibility of flattening the aggregate demand curve formed by the aggregation of more demand patterns

[28]. In a way consistent with PBR concepts, a given amount of the exceeding quantity (e.g., 5%) could be tolerated without being penalised, especially if the total time observed is relatively long (e.g., one day or more).

*5.2. Example of Duration-of-Use limits*

Let us consider again the consumer analysed in Section 4.1. The total energy below the DoU limits is 36.42 kWh, much higher than the 7.52 kWh consumed during the day, so there is room for applying flexibility to shift the consumption without any curtailment. In the best case, all the demand duration curve will remain below the DoU limits, otherwise some penalties could be possible.

The EDM outcomes obtained with the thresholds set to $\delta_1$ = 120 W and $\delta_2$ = 500 Ws are used to show an example of application of the DoU limits (Figure 5). The data referring to the DoU limits are $T_{d1}$ = 600 s (10 min), $T_{d2}$ = 1800 s (30 min), and the power values $P_{d1}$ = 3000 W, $P_{d2}$ = 2000 W, and $P_{d0}$ = 1500 W.

The excess energy with respect to the DoU limit in the first 10 min is 54 Wh, in the successive 20 min is 218 Wh, and in the remaining part of the day is 502 Wh. The application of the actions to improve flexibility needs a detailed analysis of the appliances, lifestyle of the occupants, possible home automation to control the demand, and is outside the scope of this paper.

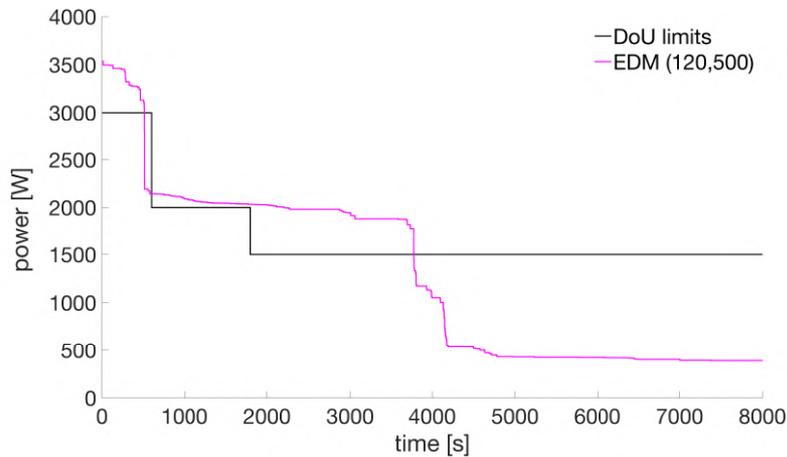

Figure 5. Example of application of DoU limits with a demand duration curve obtained from EDM outcomes.

## 6. Conclusions

In the official documents at the regulatory level, smart metering is seen as a fundamental enabler for demand side flexibility. However, all reasoning concerning the benefits, barriers and potential of smart metering is carried out with the hypothesis to exploit interval meters. The current roll-out of smart meters is expected to increase the number of devices installed with metering capabilities corresponding to time steps ranging from 15 minutes to one hour. Minute-by-minute metering is envisioned as a very effective solution, but is practically not considered at the grid connection scale.

This paper has shown how an innovative type of meter, based on measuring the energy demand between triggered events, may open new possibilities to flexibility assessment. In particular, specific innovative options come from the more precise demand pattern reconstruction obtained from event-driven energy metering. The most remarkable one is the extension of the present peak pricing to include real-time and Duration-of-Use tariff options, which take into account flexibility in a more effective way, in line with performance-based standards used in other sectors (service interruptions and power quality), and never used in the energy sector because of lack of detailed information about the consumption patterns. The detailed representation of the demand patterns enables the identification of the actual demand peaks in a way that could have been obtained only from smart meters with minute-based gathering of the metered data. However, the event-driven energy metering may provide effective demand pattern reconstruction by using a number of data lower than for minute-based interval metering. This reconstruction adds value to the knowledge of the processes that lead to the electrical demand. Establishing a monetary amount to quantify this value is challenging, also because there is no true reference case for direct comparison.

This paper has also provided a conceptualisation of the principles that may lead to the formation of a Duration-of-Use pricing scheme. An application case has been used to provide numerical evidence of a number of advantages of event-based energy metering, including the provision of more accurate data for better quantification of the network losses. From the results obtained, the event-driven energy meter is an enabler to pass from traditional calculations of quantities linked to the demand side to performance-based quantities that may consider detailed and also per cent thresholds.

The significant increase of the precision with which the technical losses in the supply grid can be assessed is a major point for the electricity distributors. In fact, with the traditional timer-driven metering with 15-minutes to 1-hour time step the losses that can be estimated when the demand has large variations, as it happens in the last mile of the electricity distribution, are largely under-estimated. This implies a twofold effect: (i) the lack of knowledge of the correct value of technical losses may misleadingly lead to attribute the remaining portion of the losses to other non-technical contributions; (ii) the allocation of the technical losses to the different customers of services could be largely inaccurate.

A similar reasoning based on Duration-of-Use principles may be used to construct a Duration-of-Use baseline and checking the demand duration against this baseline. Possible incentives and penalties are then applicable by comparing the actual demand duration with the Duration-of-Use baseline. The detailed analysis of these aspects is outside the scope of this paper. Likewise, the definition of the type of incentive or penalty scheme, and the numerical values of the incentives or penalties to be applied, have to come from a dedicated analysis carried out on the various types of consumers. Extensive testing is needed for this purpose. This matter is the subject of the present authors' work.